\documentstyle[12pt,epsbox]{article}
\begin{document}
\pagestyle{empty}
\newcommand{\hf}{\hfill}
\baselineskip=0.212in

\begin{flushleft}
\large
{SAGA-HE-94-95
\hfill October 14, 1995}  \\
\end{flushleft}

\vspace{3.0cm}

\begin{center}

\LARGE{{\bf Studies of valence-quark shadowing}} \\

\vspace{0.3cm}

\LARGE{{\bf at HERA}} \\

\vspace{2.0cm}

\Large
{S. Kumano $^*$}   \\

\vspace{1.0cm}

\Large
{Department of Physics}         \\

\vspace{0.1cm}

\Large
{Saga University}      \\

\vspace{0.1cm}

\Large
{Saga 840, Japan} \\

\vspace{2.2cm}

{proposal in ``Future Physics at HERA''} \\

\end{center}

\vspace{1.3cm}
\vfill

\noindent
{\rule{6.cm}{0.1mm}} \\

\vspace{-0.4cm}

\noindent
\vspace{-0.2cm}
\normalsize
{* Email: kumanos@cc.saga-u.ac.jp}  \\

\vspace{-0.2cm}
\noindent
\normalsize
{Research activities are listed
 at ftp://ftp.cc.saga-u.ac.jp/pub/paper/riko/quantum1} \\
\vspace{-0.6cm}

\noindent
{or at
http://www.cc.saga-u.ac.jp/saga-u/riko/physics/quantum1/kumano.html.} \\

\vfill\eject

\pagestyle{plain}
\begin{center}

\large
{Studies of valence-quark shadowing at HERA} \\

\vspace{0.5cm}

{S. Kumano} \\

{Department of Physics}  \\

{Saga University}  \\

{Saga 840, Japan} \\

\end{center}

\vspace{0.6cm}


Nuclear shadowing in the structure function
$F_2$ has been well studied \cite{SKF2}.
Models for explaining the shadowing include vector-meson-dominance-type
models and parton-recombination-type ones.
The former models describe the shadowing by the transformation of
a virtual photon into vector-meson states or $q\bar q$ states,
which then interact with a target nucleus.
The central constituents are ``shadowed" due to the existence
of nuclear surface constituents.
The latter models explain the shadowing by interactions
of partons from different nucleons in a nucleus.
They become important especially at small $x$, where
the longitudinal localization size of a parton exceeds
the average nucleon separation in the nucleus.

Because these different models produce similar $x$ and $Q^2$ dependence
in the structure function $F_2$, we cannot distinguish
among the models in comparison with experimental data.
Various shadowing models may be
tested by other quantities such as sea-quark and gluon distributions
in nuclei \cite{SKGLUE}. However, valence-quark
distributions could be useful in determining the appropriate
shadowing description \cite{KKM}.

We propose to measure the valence-quark shadowing by observing charged
pion productions in electron-nucleus scattering at HERA. In fact,
charged hadron productions have been used for finding
the u-valence-quark distribution \cite{EMC}. However, there is no
accurate data in discussing the shadowing at this stage.
In order to illustrate theoretical issues,
two different models are employed.

The first one is a hybrid parton model with $Q^2$ rescaling
and parton recombination effects \cite{SKF2}. According to the $Q^2$
rescaling model, nuclear valence-quark distribution $V_A(x,Q^2)$
is given by rescaling (increasing) $Q^2$ in the nucleon distribution
$V_N (x,Q^2)$. Therefore, the ratio $R_V\equiv V_A(x)/V_D(x)$
is smaller than unity at medium $x$ as it explains the EMC effect in
this region. Since the rescaling satisfies the baryon-number conservation
$\displaystyle{\int dx V(x) =3}$, the ratio $R_V$ becomes larger than
unity at small $x$.
Parton-recombination contributions are rather contrary to
those in the rescaling model in the sense that
the recombinations decrease the ratio at small $x$ and
increase it at medium and large $x$.
The overall effects are shown by a solid line (model 1) in Fig. 1.
In this parton model, the valence-quark shadowing differs
distinctively from the $F_2$ one:
$V_A(x) / V_N(x) \ne F_2^A(x) / F_2^N(x)$
at small $x$.

The second model is the aligned jet model in Ref. \cite{AJMF3}.
The model prediction for the valence shadowing is shown
by a solid curve (model 2) in Fig. 1.
This shadowing is very similar to the $F_2$ shadowing:
$V_A(x) / V_N(x) \approx F_2^A(x) / F_2^N(x)$
at small $x$.
It is because the $q\bar q$ pair interacts with sea and valence
quarks in the similar way.
The model curve is obtained by the aligned-jet-model
together with the baryon-number conservation.

\vspace{0.10cm}
\noindent
\parbox{7.5cm}
{
\hspace{0.6cm}
\baselineskip=0.212in
There is little experimental information on the valence shadowing
from neutrino data, so that we estimate a current experimental
restriction on the valence shadowing by using
$F_2^A/F_2^D \ (\equiv R_2)$ data at medium $x$ and the baryon
number conservation.
In finding the restriction, we assume $R_V=R_2$
in the region $x \ge 0.3$ because
valence-quark distributions dominate the $F_2$ structure functions.
We employ SLAC $R_2^{Ca}$ data, which are fitted by a smooth curve.
This curve is extrapolated into the small $x$ region
by using the baryon-number conservation.
We find that the shaded area is roughly the possible
nuclear modification, which is allowed by present experimental
data of $R_2$.
 } \ \hspace{0.5cm} \
\parbox{7.5cm}
{
\vspace{0.5cm}
\epsfile{file=94-fig1.eps,width=6.5cm}
\par\vspace{0.2cm}\hspace{0.5cm}
Fig. 1  ~Valence-quark ``shadowing".
}
\vspace{-0.1cm}

The models predict completely different
behavior at small $x$: antishadowing in the first parton model
and shadowing in the aligned-jet model.
So the models are two extreme cases, which are both
acceptable in our present knowledge as shown in Fig. 1.
We have not investigated the details of other model predictions.
However, it is very encouraging to investigate
the (anti)shadowing phenomena of the valence-quark
distribution in the sense that
the observable could be useful in discriminating among
various models, which produce similar results in
the $F_2$ shadowing.



\end{document}